# The possibility of two-exciton pairing in singlet and in triplet states

Fajian Shi
Material Sciences Division, Lawrence Berkeley National Laboratory,
One Cyclotron RD, Berkeley, CA 94720

**Abstract:** *We propose a two-dimensional exciton model showing the possibility of two-exciton pairing in cuprates, both in singlet and in triplet states. The former is assumed to be responsible for the superconductivity and the latter for the unusual normal state properties. The ground state is found to be a projection of the Bose condensation of N/2 mobile bosons with a $\eta$ projection operator which is a generalization of the Gutzwiller projection operator. The phase diagram of the model shows insulator-superconductivity and superconductivity-metal transitions with doping.*

It has been more than one decade since the discovery of high-Tc cuprate superconductors. The core mechanism of Cooper pairing in those materials that is essential to superconductivity is still unclear. The coherent length for the cuprate superconductors is extremely short (usually about $\xi \approx 14 \text{Å}$), orders smaller than that of conventional superconductors ($\xi \approx 10000 \text{Å}$). The Coulomb repulsion in such a short coherent length is very strong, so it is unlikely to have an attractive interaction between two holes. However, it was realized that an attractive interaction is not the only possible way to the formation of Cooper pairs. Anderson [1] first showed that the resonating valence-bond (RVB) states for a strongly correlated system are related to an appropriate projection of the BCS function. It is believed that the Hubbard model or t-J model should be enough to describe the high-Tc superconductivity[2]. The exchange energy J=t$^2$/U~0.4ev (t is hoping energy and U is on-site Coulomb interaction) required to break a valence-bonded pair, however, leads to a very high mean-field temperature $T_{MF}$~1000K>>Tc. It indicates at least that those models may be insufficient to describe the low energy physics much below the $T_{MF}$. Recently, Zhang [3] proposed a phenomenological theory about the SO(5) symmetry of superconductivity and antiferromagnetism. It is actually a generalization of the SO(4) symmetry for the Hubbard model [4]. New operators $\pi$ and $\pi^+$ symmetry operators carrying charge 2 and spin 1 are introduced, relating to triplet states by definition. An extended Hubbard model with the SO(5) symmetry can be constructed when the complicated infinite long-range interaction terms are introduced [5]. It is plausible that the triplet states and the long-range Coulomb correlation might be relevant to the superconductivity in cuprates. In this paper we propose a two-dimensional exciton model that includes explicitly the Coulomb interaction among acceptors and doping holes. The assumption of two-dimensionality bases on the experiments that show the incoherent transport of holes in the z-axis direction in the normal state [6] and even in the superconducting state [7].

The most important structure element of the high-Tc superconductors (both 214 compounds and 123 compounds) is the Cu-O planes. The metal or rare earth ions are sandwiched by those planes. Doping generates holes in the CuO planes and acceptors in the inter-spacing of two CuO planes. Both quantum chemistry calculations and experiments find that the doping holes have 85% O 2p and 15% Cu 3d character. Here we distinguish the doping holes and the d-holes $3d^9$ for the $Cu^{2+}$ ions. The latter exist even without any doping (hereafter holes refer to the doping holes). We use a continuous representation of Hamiltonian instead of the tight-binding (TB) models in advantage of taking account of large-range Coulomb correlation. The Hamiltonian for a system of doping holes may be written as

$$H(N,M) = \sum_{i=1}^{N} \frac{1}{2m^*} \mathbf{P}_i^2 + \sum_{i=1}^{N}\sum_{I=1}^{M} V_I(r_i) + \frac{1}{2}\sum_{i\neq j} \frac{1}{\varepsilon r_{ij}}, \tag{1}$$

where $N$ is the number of holes and $M$ is the number of acceptors $(N \leq M)$. The first term on the right side is kinetic energy corresponding to the hoping term in the TB models, $m^*$ is the effective mass of holes (which may depend on momentum as well). The second term is the interaction between holes and acceptors $V_I(r) = -1/\sqrt{d^2 + (\mathbf{r} - \mathbf{R}_I)^2}$, where $d$ is the distance from acceptor to CuO plane, $\mathbf{R}_I$ is the position in CuO plane projected from acceptor $I$. The third term is the Coulomb interaction between holes, in which $\varepsilon$ is an effective dielectric constant in CuO plane. The d-holes are not included in the Hamiltonian above, but they are indirectly relating to the effective mass and the effective dielectric constant. Notice that those holes are confined in the two-dimensional CuO plane.

It is easy to obtain the solution for a one-hole system of Hamiltonian $H(1,1)$. It is known that in a two-dimensional space there is always a bound state in a potential well no matter how shallow it is[8]. Therefore, this hole is in a bound state. A numerical solution is carried out to solve the problem. We choose a linear variational approach [9]. The trial wave function may be written as $\Phi = \sum_{i}^{K} c_i \phi_i$ where $\{\phi_i\}$ are a set of basis functions. The energy expectation value is $E = \dfrac{\int \Phi^* H_0 \Phi d\tau}{\int \Phi^* \Phi d\tau}$, where $H_0$ is $H(1,1)$. Both the eigen wave function and the eigen value of energy may be obtained from equation $\det(\mathbf{H}_0 - E\mathbf{S}) = 0$, where $\mathbf{H}_0$ and $\mathbf{S}$ are respectively the $K \times K$ matrix of the one-hole Hamiltonian and the overlap of basis functions. In our calculation, a basis set with three Gaussian functions $\phi_i = \exp(-\alpha_i r^2)$, $i = 1,3$ is chosen, and the distance $d$ is set as 4.0 atomic unit (au.). The exponents $\alpha_i$ may be optimized to give the lowest eigen value $E$. The results are listed in Table 1, in which we include also the solution for a hydrogen atom under the same approach for comparison. The coefficient $c_3$ is two orders smaller than $c_1$ and $c_2$, indicating a good convergence achieved by three Gaussian basis functions. The expectation value of radius $<r>$ for the hole is very large due to the wide half-width of potential well. We may call the bound state of the hole as a planar exciton.

The two-hole system of Hamiltonian $H(N = 2, M = 2)$ may be solved with the help of the solution for the one-hole system. The motion of two holes is correlated through the long-range Coulomb interaction $1/\varepsilon r_{12}$. This system has two spin states: singlet and triplet states. The symmetrical space wave function of the singlet state may be written as

$$\Psi^S_{AB}(1,2) = \Phi_A(1)\Phi_B(2) + \Phi_A(2)\Phi_B(1) + \lambda[\Phi_A(1)\Phi_A(2) + \Phi_B(1)\Phi_B(2)] \qquad (2)$$

where $\Phi_A$ and $\Phi_B$ are the wave functions of the excitons locating at A site and B site, $\lambda$ is a parameter in the range $[-1,+1]$. The expectation value of energy is $E_S = <\Psi^S_{AB}|H(2,2)|\Psi^S_{AB}>/<\Psi^S_{AB}|\Psi^S_{AB}>$. The parameter $\lambda_0$ for the lowest energy state of the singlet state may be determined in a variational approach $\frac{\partial E_S}{\partial \lambda} = 0$. Those wave functions in (2) with $\lambda \neq \lambda_0$ may represent excitation states. The Gaussian basis functions should be optimized again for the two-hole system. The antisymmetrical space wave function of the triplet state may be written as

$$\Psi^T_{AB}(1,2) = \Phi_A(1)\Phi_B(2) - \Phi_A(2)\Phi_B(1) \qquad (3)$$

with similar definitions as in (2). The expectation value of energy is $E_T = <\Psi^T_{AB}|H(1,2)|\Psi^T_{AB}>/<\Psi^T_{AB}|\Psi^T_{AB}>$. The Gaussian basis set may be optimized again for the triplet state. The lowest expectation value of energy for both the singlet and the triplet states depend on the distance $R_{AB} = R_A - R_B$ as well as the dielectric constant $\varepsilon$. The $R_{AB}$ dependence of $E_S$ and $E_T$ is shown in Figure (1) for a given value of the dielectric constant $\varepsilon = 1$. Notice that both the singlet and the triplet states are stable and the triplet state is the ground state for $R_{AB} \geq 3.5$ (au.). However, our calculation also shows that in the limit $\varepsilon \to \infty$ the singlet state is the ground state. The reason why the ground state depends heavily on the dielectric constant is that there is a strong competition between the exchange energy favoring the triplet state and kinetic energy favoring the singlet state. The energy gap $\Delta = E_T - E_S$ as a function of $\varepsilon$ is calculated with $R_{AB} = 10$ (au.) as shown in Figure (2). The gap $\Delta$ evolves on a continuum from negative value to positive value when the dielectric constant increases from one to infinity. For a set of parameters $\varepsilon = 2.5$, $m^* = m_e$, and $R_{AB} = 10$ (au.), the magnitude of gap is obtained about $\Delta = 50$ mev the same magnitude of gap for high-Tc superconductors. Moreover, the gap is disappearing for the distance $R_{AB} > 18$ (au.) as shown in Figure (1). The diameter of two excitons with $R_{AB} = 18$ (au.) is about $36$ (au.) $\approx 18$ Å. This length scale is also the same as the coherent length in high-Tc superconductors. Therefore, the simple exciton model is able to give the right magnitude of energy gap and the coherent length, assuming that the superconducting gap be the gap between the triplet state and the singlet state.

The many body wave functions for a system of N holes described by Hamiltonian H(N,M) with N=M>2 may be expanded in terms of the one and two-hole wave functions. For instance, let's consider a simple system of four holes. Four holes may form two two-exciton pairs, in the singlet or the triplet state. The number of configurations is 3 for two-exciton pairing in the singlet state, 3 in the triplet state and 6 in a mix of the singlet state and the triplet state. If the 3-fold degeneracy of triplet pairing is considered, there are 48 possible quantum states, 48=3*1+3*9+6*3. The Hamiltonian H(4,4) may be solved in the space spanned by the 48 basic functions. The eigen states and eigen values may be obtained from a $48 \times 48$ matrix of the Hamiltonian. As we have shown before, the energy level of single exciton split into two energy levels corresponding to the triplet and singlet state with a finite gap when two excitons get closer. For a system of many excitons, the two energy levels expands further into two energy bands due to the overlapping of wave functions. If the gap is larger than the band width, the ground state may be the lower band with two excitons pairing in the singlet state. For the simple system of 4-excitons, the wave function of the ground state may be expressed as a combination of the three configurations with each pair in singlet state. The above picture may be generalized to an N-hole system of H(N,N). The wave function of each configuration may take the form

$$|\Psi_{\{IJ,I'J'\}}\rangle = \prod_{\{IJ\}}{}'|\Phi^S{}_{IJ}\rangle \prod_{\{I'J'\}}{}'|\Phi^T{}_{I'J'}\rangle, \qquad (4)$$

with $|\Phi^S{}_{IJ}\rangle = |\Psi^S_{IJ}(1,2)\rangle|\chi_S\rangle$ and $|\Phi^T{}_{I'J'}\rangle = |\Psi^T_{I'J'}(1,2)\rangle|\chi_T\rangle$. The ' in the product $\prod'$ indicates no overlapping among the sites $\{IJ, I'J'\}$. The number of possible configurations is proportional to $N!$. The eigen states may be expressed generally as the combination of all the configurations, $|\Psi_E\rangle = \sum_{\{IJ,I'J'\}} \alpha_{\{IJ,I'J'\}}|\Psi_{\{IJ,I'J'\}}\rangle$. Again for the large gap between the triplet state and the singlet states, the lowest eigen states may be written as the combination of those configurations with those two-exciton pairs in singlet states $|\Psi_G\rangle = \sum_{\{IJ\}} \alpha_{\{IJ\}}|\Psi_{\{IJ\}}\rangle$.

Since only those two-exciton pairs at neighbor sites are important for the ground state, the $|\Psi_G\rangle$ may be reduced to $|\Psi_G\rangle = \sum_{\{<IJ>\}} \alpha_{\{<IJ>\}}|\Psi_{\{<IJ>\}}\rangle$, where $<IJ>$ indicates the neighbor sites. Furthermore, the coefficients $\alpha_{\{<IJ>\}}$ for those configurations $\{<IJ>\}$ may be assumed to be the same, because those configurations have almost the same energy or are highly "degenerate". The singlet state of a two-exciton pair may be written of form $|\Phi^S{}_{IJ}\rangle = [c^+_{I\uparrow}c^+_{J\downarrow} + c^+_{I\downarrow}c^+_{J\uparrow} + \lambda(c^+_{I\uparrow}c^+_{I\downarrow} + c^+_{J\uparrow}c^+_{J\downarrow})]|0\rangle$, where $c^+_{I\sigma}$ and $c^+_{I\sigma}$ are the single-exciton creation operators. The ground state $|\Psi_G\rangle$ is then expanded

$$|\Psi_G\rangle = \sum_{\{<IJ>\}} \prod_{\{<IJ>\}}{}'|\Phi^S_{IJ}\rangle = A + \lambda B + \lambda^2 C + \cdots, \qquad (5)$$

with $A = \sum (c^+_{I_1\sigma_1} c^+_{I_2\sigma_2} \cdots c^+_{I_k\sigma_k} \cdots c^+_{I_N\sigma_N})|0\rangle$ summing over configurations without any double occupation, $B$ with only one double occupation and one vacancy, and $C$ with only two occupations and two vacancies and so on. We may compare this ground state with a Bose condensation of mobile bosons. We use the representation given by Anderson[1]. A mobile boson may be written as

$$b^+ = \sum_k a(k) c^+_{k\uparrow} c^+_{-k\downarrow} \cong \frac{1}{\sqrt{M}} \sum_{\tau=(nn)} \sum_J c^+_{J\uparrow} c^+_{J+\tau\downarrow}|0\rangle = \frac{1}{\sqrt{M}} \sum_{<IJ>} c^+_{I\uparrow} c^+_{J\downarrow}|0\rangle, \qquad (6)$$

where $nn$ indicates nearest neighbors. We can make a Bose condensation of N/2 bosons by forming

$$|\Psi_{BS}\rangle = (b^+)^{N/2}|0\rangle = \frac{1}{N^{N/2}}(\sum_{<IJ>} c^+_{I\uparrow}c^+_{J\downarrow})^{N/2}|0\rangle = \frac{1}{N^{N/2}}\sum_{\{<IJ>\}}\prod_{\{<IJ>\}} c^+_{I\uparrow}c^+_{J\downarrow}|0\rangle = A+B+C+\cdots. \quad (7)$$

which is equivalent to the ground state $|\Psi_G(\lambda=1)\rangle$. Obviously, $\lambda=1$ is related to a vanishing Coulomb interaction in Equation (1), i.e. $\varepsilon \to \infty$ The ground state of Hamiltonian (1) without Coulomb interaction is a superconducting state! Usually the parameter $\lambda$ is smaller than 1 for finite Coulomb interaction. We introduce a $\eta$ projection operation

$$P_d^\eta = \prod_I (1 - \eta n_{I\uparrow} n_{I\downarrow}) \quad (8)$$

where $n_I$ is the occupation number operator. Acting the projection operator (8) on the state of Bose condensation (7) leads to a new state: $P_d^\eta \Psi_{BS} = A + (1-\eta)B + (1-\eta)^2 C + \cdots$. The ground state is thus equivalent to the $\eta$ projection of the Bose condensation,

$$\Psi_G = P_d^\eta \Psi_{BS} \quad (9)$$

with $\eta = 1 - \lambda$. In the strong correlation limit we have $\lambda = 0$ or $\eta = 1$. The ground state becomes the RVB insulator defined by Anderson[1]. Moreover, $\Psi_{BS}$ is a projection of the BCS function $\Psi_{BS} = P_{N/2}\Psi_{BCS}$ where $P_{N/2}$ is a projection operator onto N/2 pairs. The ground state is thus an appropriate projection of the BCS function $\Psi_G = P_d^\eta P_{N/2} \Psi_{BCS}$. The $\eta$ projection operator looks like a kind of "mobility" operator that confines the mobility of a Bose condensation state $\Psi_{BS}$. It has to emphasize that the ground state we assume is stable only if the gap between the triplet and singlet states is significant. In addition, the number of holes on CuO may be less than the number of acceptors, $N < M$ when the concentration of oxygen changes in cuprates. In this case even the RVB state might become superconducting [1]. The parameter $\lambda$ is becoming negative when two excitons are too close, corresponding to $\eta > 1$. The projection of Bose condensation becomes $P_d^\eta|\Psi_{BS}\rangle = A + |1-\eta|e^{i\pi}B + |1-\eta|^2 e^{i2\pi}B + \cdots$. It equals to add a phase factor $\exp(n\pi)$ to the states with the n-double occupations. Further one may introduce a non-integral phase in the projection operator when the density of holes is very low and phase fluctuations are important [10]. Because the high-Tc cuprates are in metallic phase in the high doping region, the projected state (9) for $\eta > 1$ might describe a metal phase. The wave function $|\Psi_G\rangle$ might represent either a insulator or a superconductor or a metal. It has to emphasize that the deduction of the equations from (5) to (9) does not depend on the dimensionality of space. The wave functions $\Psi_G = P_d^\eta \Psi_{BS}$ may be considered as the basis functions for most of superconductivity. The BCS function is indeed an expansion in terms of those these basis functions with $\lambda = 1$ and variable number of pairs. In the case of $\lambda \neq \lambda_0$, the Equation (9) represents the low-lying excitation states that are obviously still the projection of Bose condensation. Therefore the superconductivity is stable at finite temperature. In addition, due to that continuous spectrum, the specific heat should be a power function of temperature, rather than the exponential function for metal superconductors[11] at low temperature.

The phase diagram of H(N,N) depends on those parameters: temperature $T$, dielectric constant $\varepsilon$ and average distance between holes $R_{av}$. The last one relates to the density of holes in terms of $n = (R_0 / R_{av})^2$ where $R_0$ is the lattice constant for the square lattice of the CuO plane. In cuprates, the dielectric constant should also depend on the density of holes. Therefore the phase diagram depends mainly on hole density as well as temperature. In the low doping limit, the gap between two spin states is vanishingly small. The Hamiltonian H(N,N) may be mapped to a large-U Hubbard model with $U \gg t$ at half-filling, $H_U = \sum_{IJ\sigma} t_{IJ} c_{I\sigma}^+ c_{J\sigma} + U \sum_I n_{I\uparrow} n_{Ii\downarrow}$, where $c_{I\sigma}^+$ ($c_{J\sigma}$) is the creation (annihilation) operator for the bound states of excitons. The ground state of the model is an antiferromagnetic insulator or a RVB insulator. In intermediate doping region, the gap between the singlet and triplet states is significantly large. The doping dependence of the energy gap is obtained by $\Delta(n) = \Delta(R)$ with $R = R_0 (1/n)^{1/2}$, $R_0 = 3.5$ au. and $\varepsilon = 2.5$, as shown in Figure (3). The gap is obviously nonzero and positive in the region $0.05 < n < 0.45$ where the ground state should be a superconducting state. Above Tc, the two-exciton pairs in triplet state are dominating, which may be responsible for the unusual normal phase properties. In even higher doping region, because the potential wells overlap with each other strongly, the second term of the Hamiltonian (1) is nearly a constant. The exciton description of holes is no longer valid. The Hamiltonian then reduces to a two-dimensional interacting Fermi gas that we believe should be metallic. In conclusion, the Hamiltonian $H(N,N)$ has three phases: the AF or RVB insulator phase in low doping limit, superconducting phase in intermediate doping region; and metal phase in the higher doping region as shown in Figure (3). In the low doping limit of our phase diagram, we do not find the so-called pseudogap [12] which has a magnitude of 1000KT. The pseudogap might be contributed by the d-holes because it has the same magnitude as exchange energy $J$. The fact that both the short-range antiferromagnetic order and the superconductivity coexist suggests that the d-holes might be responsible for the short-range antiferromagnetic order and the doping holes for the superconductivity.

The triplet excitation states may contribute to many unusual properties of high-Tc superconductors. First, the probability of a singlet state being excited is proportional to $3\exp(-\Delta/kT)$ instead of $\exp(-\Delta/kT)$ as in conventional superconductors. This may explain the large ratio $\frac{\Delta}{kT_c}$ observed in cuprate superconductors. In addition, the extra number of states should be detected in experiments when a two-exciton pair is excited from a singlet state into a triplet state. Second, the triplet states may contribute to an enhanced susceptibility of the normal phase above Tc [13]. Furthermore, the transport properties in normal phase may be different from the Landau Fermi liquid, since the triplet states are bosons. It would be interesting to study the thermodynamics and magnetic properties of such a system at finite temperature. We leave them in future work.

In conclusion, the system of doping holes in cuprate superconductors is described by a two-dimensional exciton model that includes the long-range Coulomb correlation. Two holes may form a pair in singlet state and in triplet state, the former is responsible for the superconducting state and the latter is responsible for the unusual normal state properties. The ground state is found to be the projection of a Bose condensation that may be either a RVB insulator, or a superconductor or a metal. The $\eta$ projection operator introduced is actually a generalization of the Gutzwiller projection operator. It is the two-dimensionality not the Coulomb correlation that is the driving force of superconductivity. The inclusion of long-range Coulomb correlation leads to a phase diagram that shows the insulator-superconductor transition as well as superconductor-metal phase transition. Our exciton model is very different from the exciton mechanism [14] that the exchange of exciton between two electrons is assumed to responsible for a net attractive force in analogy to the BCS theory.

**Acknowledgement:** The author thanks Dr. Michel Van Hove for discussions and encouragement.

**Figure Captions**

**Figure 1.** Energy expectation value $E_S$ and $E_T$ of two-exciton pairing in singlet and triplet states versus the distance of the two excitons $R_{AB}$. Dielectric constant is set as $\varepsilon = 1$. One energy unit equals to 27.2 ev.

**Figure 2.** Energy gap $\Delta = (E_T - E_S)$ between the triplet and the singlet states versus dielectric constant $\varepsilon$. The distance between two excitons is set as $R_{AB} = 10$ **au.**.

**Figure 3.** Phase diagram of the Hamiltonian (1) and the doping dependence of gap for a given dielectric constant $\varepsilon = 2.5$ and lattice constant $R_0 = 3.5$ **au.**.

**Table 1. The results of numerical calculation for a single exciton and a Hydrogen atom in the variational approach.**

**Wave function $c_i$ (unrenormalized), optimized Gaussians $\alpha_i$, expectation value of energy $E$ and mean radius $<r>$ (in atomic unit) for exciton and Hydrogen atom are listed.**

| $c_i$ (exciton) | $\alpha_i$ (exciton) | $E$ (Hartrees) / $<r>$ (au.) | $c_i$ (Hydrogen) | $\alpha_i$ (Hydrogen) | $E$ (Hartrees) / $<r>$ (au.) |
|---|---|---|---|---|---|
| $c_1 = 1.0$ | $\alpha_1 = 0.0198$ | $-0.159$ | $c_1 = 1.0$ | $\alpha_1 = 0.159$ | $-0.497$ |
| $c_2 = 1.04$ | $\alpha_2 = 0.0582$ | $/ 7.228$ | $c_2 = 1.843$ | $\alpha_2 = 0.775$ | $/ 1$ |
| $c_3 = 0.0056$ | $\alpha_3 = 3.3281$ | | $c_3 = 1.168$ | $\alpha_3 = 7.54$ | |

Figure 1

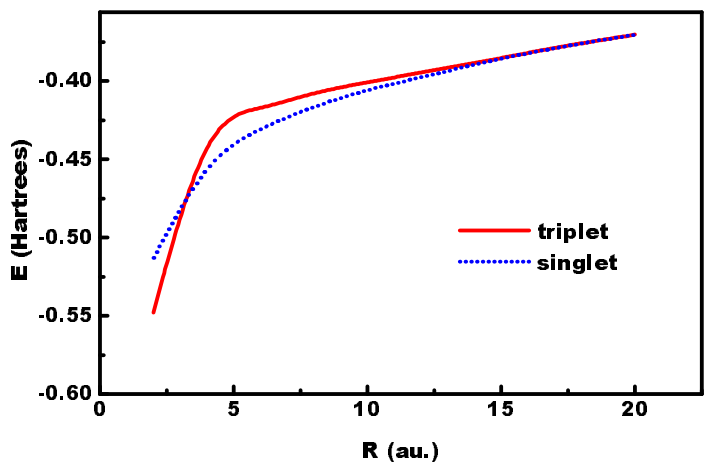

**Figure 2**

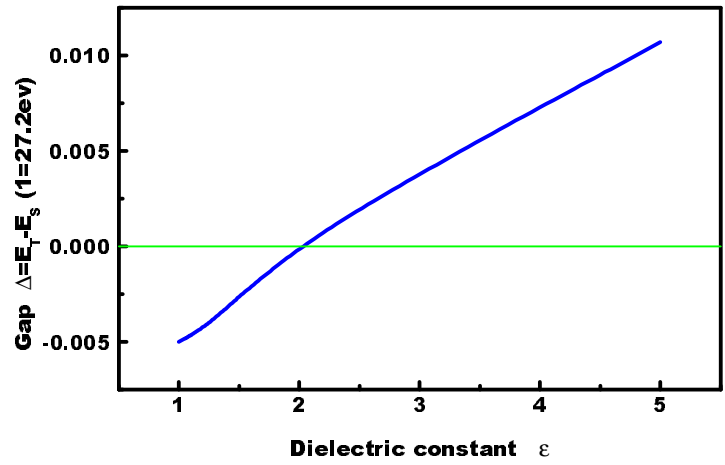

**Figure 3**

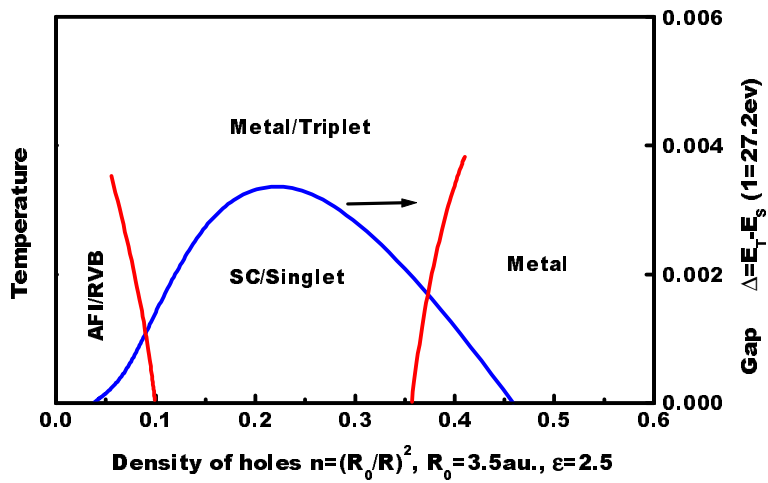